\pdfoutput=1
\documentclass[11pt,twoside,a4paper,cmspaper,final,collab]{cms-tdr}
\def\svnVersion{7fba794}\def\svnDate{2025/09/10}\def\cmsCernNoTag{CERN-EP-2025-030}\def\cmsCernDate{\today}\def\cmsMessage{Published in Physical Review Letters as \href{http://dx.doi.org/10.1103/w9kp-f8xr}{\doi{10.1103/w9kp-f8xr}.}}
\begin{document}\cmsNoteHeader{HIN-23-009}

\newlength\cmsTabSkip\setlength{\cmsTabSkip}{1ex}

\newcommand{\mumu}{\ensuremath{\PGmp\PGmm}\xspace}
\newcommand{\WgN}  {\ensuremath{W_{\PGg{}\mathrm{N}}}\xspace}
\newcommand{\RglPb} {\ensuremath{R^{\mathrm{Pb}}_{\Pg{}}}\xspace}
\newcommand{\Sjpsi} {\ensuremath{\mathrm{S^{\PJGy}}}\xspace}
\newcommand{\nGama} {\ensuremath{n_{\PGg}}\xspace}
\newcommand{\PbPb}  {\ensuremath{\text{PbPb}}\xspace}
\newcommand{\pb}  {\ensuremath{\text{Pb}}\xspace}
\newcommand{\STARlight}{\textsc{STARlight}\xspace}

\title{Probing gluon fluctuations in nuclei with the first energy-dependent measurement of incoherent \texorpdfstring{$\PJGy$}{J/psi} photoproduction in ultraperipheral PbPb collisions}

\date{\today}

\abstract{
Incoherent $\PJGy$ photoproduction in heavy ion ultraperipheral collisions (UPCs) provides a sensitive probe of localized, fluctuating gluonic structures within heavy nuclei.
This study reports the first measurement of the photon-nucleon center-of-mass energy (\WgN) dependence of this process in \PbPb UPCs at a nucleon-nucleon center-of-mass energy of 5.02\TeV, using 1.52\nbinv of data recorded by the CMS experiment.
The measurement covers a wide \WgN range of $\approx$40--400\GeV, probing gluons carrying a fraction $x$ of nucleon momentum down to an unexplored regime of $6.5 \times 10^{-5}$. 
Compared to baseline predictions neglecting nuclear effects, the measured cross sections exhibit significantly greater suppression at lower $x$. 
Additionally, the ratio of incoherent to coherent photoproduction is found to be constant across the probed $\WgN$ and $x$ range, disfavoring the establishment of the black disk limit. 
This study provides critical insights into the $x$-dependent evolution of fluctuating gluonic structures within nuclei and calls for further advancements in theoretical models incorporating nuclear shadowing and gluon saturation.
}

\hypersetup{%
pdfauthor={CMS Collaboration},%
pdftitle={Probing gluon fluctuations in nuclei with the first energy-dependent measurement of the incoherent J/Psi photoproduction in ultraperipheral PbPb collisions},%
pdfsubject={CMS},%
pdfkeywords={CMS, UPC, gluon saturation, fluctuation, nuclear shadowing, photon-nuclear}}

\maketitle 

Understanding the spatial and momentum structure of nuclei at the most fundamental level is a central goal of quantum chromodynamics (QCD), the theory of the strong interaction. The structure of nucleons and nuclei becomes increasingly dominated by gluons~\cite{H1:2015ubc,Frankfurt:2005mc} when probed at higher energies or smaller values of the momentum fraction $x$ carried by partons~\cite{Bjorken:1968dy}. Gluon densities grow rapidly at small $x$, and different physical mechanisms can modify this growth. One such effect is ``nuclear shadowing,'' which arises from the coherent interaction of a probing photon with multiple nucleons in the nucleus, leading to a suppression of the effective gluon distribution~\cite{Frankfurt:2005mc}. When gluon densities become extremely high at small $x$, a new phenomenon named ``gluon saturation'' is expected to occur because of nonlinear interactions among gluons themselves~\cite{Iancu:2003xm,Gelis:2010nm}. While both phenomena reduce gluon densities at small $x$, they reflect distinct underlying physics.

In ultraperipheral collisions (UPCs)~\cite{Baltz:2007kq}, where relativistic heavy ions interact at impact parameters exceeding the sum of their nuclear radii, strong interactions are highly suppressed. In these collisions, a quasi-real photon emitted by an ion can interact with the other ion and induce the diffractive production of a vector meson. Notably, the diffractive photoproduction of heavy vector mesons, e.g., $\PJGy$, is highly sensitive to the gluon distribution within the target nucleus, offering a powerful probe of the nuclear gluonic structure.
This production can occur through coherent or incoherent processes. In coherent photoproduction the photon interacts with the entire nucleus, which remains intact and in its ground state, providing insight into the average nuclear gluon density~\cite{Guzey:2013jaa, Klein:2020fmr}. 
In contrast, incoherent photoproduction involves interactions between the photon and localized gluonic hotspots, often leading to nuclear excitation or breakup, revealing event-by-event fluctuations in the spatial configuration of the nuclear gluon field~\cite{ALICE:2023gcs, Mantysaari:2020axf, Caldwell:2010zza}. These fluctuations are particularly relevant at small $x$ and may be connected to the initial-state geometry of heavy ion collisions and the emergence of collective phenomena in small system collisions~\cite{Dusling:2015gta, Nagle:2018nvi, Mantysaari:2020axf}.

Ultraperipheral collisions have been extensively used to study coherent and incoherent vector meson photoproduction across a wide range of collision energies, nuclear systems, and rapidities~\cite{Afanasiev:2009hy, ALICE:2014eof, STAR:2022wfe, Abelev:2012ba, ALICE:2013wjo, Acharya:2019vlb, Acharya:2019vlb, ALICE:2021gpt, Khachatryan:2016qhq, LHCb:2022ahs,CMS:2024krd,CMS:2025lsm}. 
These measurements have demonstrated strong nuclear modifications to the cross sections, providing key tests of models of gluon distributions in nuclei. 
Recent breakthrough studies by the CMS~\cite{CMS:2023snh} and ALICE~\cite{ALICE:2023jgu} Collaborations have provided the first measurements of the dependence of coherent $\PJGy$ photoproduction on the photon-nucleon center-of-mass energy ($\WgN$). 
The results reveal significantly greater nuclear suppression at higher $\WgN$ (lower $x$) compared to the impulse approximation (IA)~\cite{Guzey:2013xba,Guzey:2018tlk,Kryshen:2023bxy}, which neglects nuclear effects and treats the nucleus as a collection of free nucleons. 
However, models incorporating gluon saturation or nuclear shadowing fail to fully describe the data. The ``black disk limit" phenomenon~\cite{Frankfurt:2005mc,Frankfurt:2001nt} may offer a qualitative interpretation~\cite{CMS:2023snh}. This limit hypothesizes that the nucleus becomes a uniform, complete absorber of incoming photons at small $x$, maximizing the coherent photoproduction cross section while strongly suppressing or eliminating incoherent photoproduction.
Measurements of the energy dependence of incoherent $\PJGy$ photoproduction, particularly at the high-energy limit, provide a direct test of this phenomenon.
However, such measurements remain scarce~\cite{STAR:2023nos,ALICE:2023gcs}, primarily due to challenges in distinguishing low- and high-energy photon contributions, as detailed in later sections. These limitations have hindered studies of fluctuating gluon evolution dynamics within nuclei, especially at small $x$, for decades. Furthermore, comparing incoherent and coherent channels reduces theoretical uncertainties in the $\PJGy$ wavefunction, photon flux, nuclear density, nuclear form factor, nucleon parton distribution functions, and the $\PJGy$ formation cross section from single $\PGg{}$-nucleon scattering~\cite{Mantysaari:2017dwh,Guzey:2018tlk,Kryshen:2023bxy}. This enables stringent tests of models incorporating subnucleonic fluctuations, nuclear shadowing, and nonlinear QCD effects~\cite{Frankfurt:2005mc,Lappi:2010dd,Cepila:2016uku,Mantysaari:2016ykx,Mantysaari:2020axf}. 

This Letter presents the first measurement of the energy dependence of incoherent $\PJGy$ photoproduction in lead-lead (PbPb) UPCs at a nucleon-nucleon center-of-mass energy $\sqrtsNN = 5.02\TeV$ with an integrated luminosity of $1.52\nbinv$. 
The results span an energy range of $40<\WgN<400\GeV$, probing $x$ values as low as $6.5 \times 10^{-5}$. 
By quantifying the suppression relative to baseline calculations using the IA and studying the ratio of incoherent to coherent production, these measurements offer new constraints on the evolution of spatial structure and fluctuations of gluon fields in nuclei at small $x$. 
The data and uncertainties, along with the covariance matrix reflecting correlations among data points, are available at HEPData~\cite{hepdata}.

The CMS apparatus~\cite{CMS:2008xjf} is a multipurpose, nearly hermetic detector, designed to trigger on and identify electrons, muons, photons, and hadrons~\cite{CMS:2020cmk,Khachatryan:2016bia,CMS:2020uim,CMS:2018rym,CMS:2014pgm}. 
A global ``particle-flow" algorithm~\cite{CMS:2017yfk} reconstructs all particles in an event. It integrates information from an all-silicon tracker, a crystal electromagnetic calorimeter, and a brass and scintillator hadron calorimeter operating inside a 3.8\unit{T} superconducting solenoid, and muon detectors within the flux-return yoke.
Forward calorimeters~\cite{Bayatian:2006jz}, made of steel and quartz fibers, extend the pseudorapidity ($\eta$) coverage provided by the barrel and end-cap detectors.
Two zero-degree calorimeters (ZDCs)~\cite{Suranyi:2021ssd}, made of quartz fibers and plates embedded in tungsten absorbers, are used to detect forward neutrons.

The dimuon decay channel is utilized to reconstruct $\PJGy$ candidates.
Events are selected using a hardware-based trigger system requiring at least one muon candidate coincident with a PbPb bunch crossing, without a \pt threshold~\cite{Khachatryan:2016bia}. Events with energy deposits above the noise threshold in either forward calorimeter are vetoed at the trigger level. Offline, a primary interaction vertex is required within 20\unit{cm} along the beam axis from the center of the nominal interaction region and 2\unit{cm} from the beam axis in the transverse plane, formed from at least two tracks. To suppress hadronic interactions, forward calorimeter energy deposits must not exceed 7.3 and 7.6\GeV on the positive and negative rapidity sides, respectively~\cite{CMS:2020skx}. Events must contain exactly two oppositely charged muons with an invariant mass of $2.6 < m_{\mumu} < 4.2\GeV$ and no extra tracks within $\abs{\eta} < 2.4$. Both muons must meet the ``soft muon" criteria~\cite{Chatrchyan:2012xi}, combining tracker and muon detector information, and at least one must match the trigger. Backgrounds, including same-sign dimuon pairs, are negligible after these selections.

The raw number of $\PJGy$ signal events contains contributions from several physics processes: 
incoherent (with or without nucleon dissociation) and coherent direct $\PJGy$ production from photon-nucleus interactions, and also $\PJGy$ coming from the decay of directly photoproduced $\PGyP{2S}$, referred to as ``feed-down.''
The incoherent component is extracted by fitting the dimuon invariant mass and $\pt$ distributions after correcting for acceptance ($A$) and efficiency ($\epsilon$) effects in each rapidity bin. The fitting procedures are similar to those of~\cite{Acharya:2019vlb,ALICE:2021gpt,CMS:2023snh} (more details are available in \appendixname~\ref{appendix:a}).
The values of $(A \epsilon)_{\PJGy}$, are estimated using a simulated sample of incoherent $\PJGy \to \PGmp\PGmm$ events generated with the \STARlight (v3.13) event generator~\cite{Klein:2016yzr}. 
These events are processed with the full CMS detector response simulation using \GEANTfour~\cite{Agostinelli:2002hh}. Additional corrections determined using a tag-and-probe technique~\cite{CMS:2011aa,Chatrchyan:2012xi} are applied to account for any data-to-simulation discrepancies.

The differential cross section for incoherent $\PJGy$ photoproduction is given by:
\begin{equation}
\frac{\rd \sigma^{\text{incoh}}_{\PJGy}}{\rd y} = \frac{N^{\text{incoh}}_{\PJGy}}{(A \epsilon)_{\PJGy} \mathcal{B}_{\PJGy \to \mumu} \lumi_{\text{int}} \Delta y}.
\end{equation}
Here, $N^{\text{incoh}}_{\PJGy}/(A \epsilon)_{\PJGy}$ is the yield of $\PJGy$ produced via incoherent processes, corrected for acceptance and efficiency. 
The branching fraction $\mathcal{B}_{\PJGy \to \mumu} = (5.961 \pm 0.033)\%$ is taken from Ref.~\cite{ParticleDataGroup:2024cfk}, $\lumi_{\text{int}}$ is the total integrated luminosity~\cite{CMS:2022bjp}, and $\Delta y$ is the rapidity bin width.

In \PbPb UPCs, a $\PJGy$ produced at rapidity $y$ can result from two photon energies, $\omega_1 = (M_{\PJGy}/2)\re^{-y}$ and $\omega_2 = (M_{\PJGy}/2)\re^{+y}$, corresponding to photon emissions from opposite nuclei~\cite{CMS:2023snh}. This ambiguity affects $x$ and $\WgN$, calculated as $x = (M_{\PJGy}/\sqrtsNN)\re^{\pm y}$ and $\WgN = \sqrt{\sqrtsNN M_{\PJGy}\re^{\mp y}}$. The total cross section at rapidity $y$ combines contributions from both photon energies~\cite{Guzey:2013jaa,Guzey:2016piu}:
\ifthenelse{\boolean{cms@external}}
{
	\begin{multline}
	\label{eq:two-way-sum}
	\frac{\rd \sigma_{\pb\pb \to \pb\pb'\PJGy} (y)}{\rd y}  
	\\ = \nGama(\omega_1)\sigma_{\PGg{} \pb\to \PJGy \pb'}(\omega_1) + \nGama(\omega_2) \sigma_{\PGg{} \pb\to \PJGy \pb'}(\omega_2),
	\end{multline}
}
{
	\begin{equation}
	\label{eq:two-way-sum}
	\frac{\rd \sigma_{\pb\pb \to \pb\pb'\PJGy} (y)}{\rd y}  
	= \nGama(\omega_1)\sigma_{\PGg{} \pb\to \PJGy \pb'}(\omega_1) + \nGama(\omega_2) \sigma_{\PGg{} \pb\to \PJGy \pb'}(\omega_2),
	\end{equation}
}
where $\nGama$ is the photon flux, and $\sigma_{\PGg{} \pb\to \PJGy \pb'}$ is the cross section for $\PJGy$ production.

To resolve the ``two-way ambiguity'', neutron tagging is used based on ZDC energy deposits to classify events as 0n (no neutrons) or Xn (one or more neutrons)~\cite{CMS:2020skx, CMS:2023snh}. Events without neutron multiplicity selection are labeled as An. Combining the classifications from both ZDC sides yields three event categories, denoted 0n0n, 0nXn, and XnXn, that are used for cross section measurements. Neutron peaks in the ZDC energy distributions are fitted with a multi-Gaussian function, achieving purities above 99.6\% for Xn neutron signals.
Neutrons can be emitted from both electromagnetic dissociation (EMD) or incoherent photoproduction processes. The large EMD cross section ($\sim$200\unit{b})~\cite{Pshenichnov:2001qd} leads to concurrent neutron emission from other \PbPb interactions in the same bunch crossing, which can migrate the neutron multiplicity class to a higher one. Such EMD pileup effects are corrected using the zero-bias triggered data (requiring only that two beams cross each other), where the migration probability is directly determined by the probability of observing each neutron multiplicity class~\cite{CMS:2020skx}. 

In incoherent $\PJGy$ photoproduction, neutrons primarily align with the outgoing direction of the target nucleus~\cite{Guzey:2013jaa,Guzey:2016piu}. In 0nXn events, $\PJGy$ with the same or opposite sign rapidity as the detected neutrons correspond to contributions from $\omega_1$ and $\omega_2$, respectively. 
Calculating the associated photon flux is challenging without a robust understanding of incoherent photoproduction itself. 
However, the photon flux can be extracted using EMD-induced neutron events~\cite{Klein:2016yzr}. 
To facilitate photon flux determination, the 0n0n and 0nXn events are combined into a single event class, denoted as 0nAn*, under the assumption that the $\omega_1$ and $\omega_2$ fractions for 0n0n events are similar to those in 0nXn events. 
An alternative assumption, where 0n0n events only contribute to $\omega_1$, is used as an extreme limit to estimate systematic uncertainties.

The incoherent $\PJGy$ production cross section per $\PGg{}\pb$ interaction is calculated as:
\begin{equation}
\sigma_{\PGg{} \pb\to \PJGy \pb'}(\omega) = \frac{\rd \sigma^{\text{0nAn*}}_{\pb\pb\to \pb\pb'\PJGy} (y)}{\rd y} \: / \: \nGama^{\text{0nAn*}}(\omega),
   \label{eq:sigma_0nAn}
\end{equation}
\noindent where $\nGama^{\text{0nAn*}}(\omega)$ is the total photon flux associated with events having no neutrons on one side of the ZDC. This flux can be determined using EMD-induced contributions:
\begin{equation}
	\label{eq:flux_0nAn}
	\nGama^{\text{0nAn*}}(\omega) = \nGama^{\text{0n0n(EMD)}}(\omega) + \frac{1}{2}\nGama^{\text{0nXn(EMD)}}(\omega).
\end{equation}
\noindent where $\nGama^{\text{0n0n(EMD)}}$ and $\nGama^{\text{0nXn(EMD)}}$ represent the photon fluxes for 0n0n and 0nXn EMD events, respectively. These flux components are derived from equivalent photon approximation models implemented in \STARlight~\cite{Klein:2016yzr}. The factor of 1/2 accounts for the equal probability of EMD-induced neutrons being emitted in the same or opposite direction as incoherent-induced neutrons, with the latter case contributing to the XnXn classification instead.
Alternatively, given a 0.85 probability of an incoherent process with forward neutrons~\cite{Guzey:2013jaa}, Eq.~(\ref{eq:flux_0nAn}) can be updated by replacing 1/2 with 0.575 ($= 1-0.85/2$). Here, 0.575 represents the fraction of EMD-induced 0nXn events retained in 0nAn* events.
These two approaches differ by approximately $1\%$ and 3.7--4.7\% for the $\nGama^{\text{0nAn*}}$ values at low and high $\WgN$, respectively. These differences are incorporated into the systematic uncertainties of the cross-section measurements.

Systematic uncertainties are evaluated by taking the maximum deviation from the nominal result for each source. The uncertainty in the integrated luminosity is $1.7\%$~\cite{CMS:2021xjt,CMS:2024uor}, and the uncertainty in $\mathcal{B}_{\PJGy \to \mumu}$ is $0.55\%$~\cite{ParticleDataGroup:2024cfk}. 
Fit-related uncertainties in signal yield extraction are evaluated by varying the signal and background models, as follows: (i)~using a sum of Crystal Ball~\cite{Oreglia:1981fx} and Gaussian functions for the $\PJGy$ and $\PGyP{2S}$ signals; 
(ii)~fixing the Crystal Ball parameters to values determined from Monte Carlo simulations;
(iii)~using a fourth-order polynomial for the quantum electrodynamics (QED) background;
(iv)~adjusting the dimuon invariant mass range in the fit and sideband mass region for the QED background \pt template; 
(v)~varying the coherent-process-dominated \pt region to be $\pt<0.3\GeV$;
(vi)~varying the Pb nuclear radius by 1\unit{fm} for the coherent \PJGy $\pt$ template simulation as suggested in Ref.~\cite{CMS:2023snh};
(vii)~allowing free parameters for the \pt distributions of incoherent \PJGy and $\PGyP{2S}$ with nucleon dissociation.
They contribute a total of 6--13\%, depending on \WgN.
Tag-and-probe corrections introduce uncertainties of 3--7\%, while the choice of forward calorimeter threshold leads to a 4--8\% uncertainty. 
The uncertainty in the neutron multiplicity migration correction is 0.1--0.2\%, determined by comparing an alternative approach~\cite{ATLAS:2020epq} with the observed neutron emission rates. 
The uncertainties from fractions at low and high $\omega$ in 0n0n events are 1--7\% depending on the \WgN values.
Individual sources of experimental uncertainties are added in quadrature to obtain the total systematic uncertainty of 10--18\%. 
Uncertainties stemming from photon flux calculations are below $3\%$, reflecting variations due to uncertainties in the Pb nuclear radius, nuclear skin thickness~\cite{dEnterria:2020dwq}, and the EMD cross section~\cite{Baltz:1996as}.

\begin{figure*}[hbt!]
\centering
\includegraphics[width=0.49\textwidth]{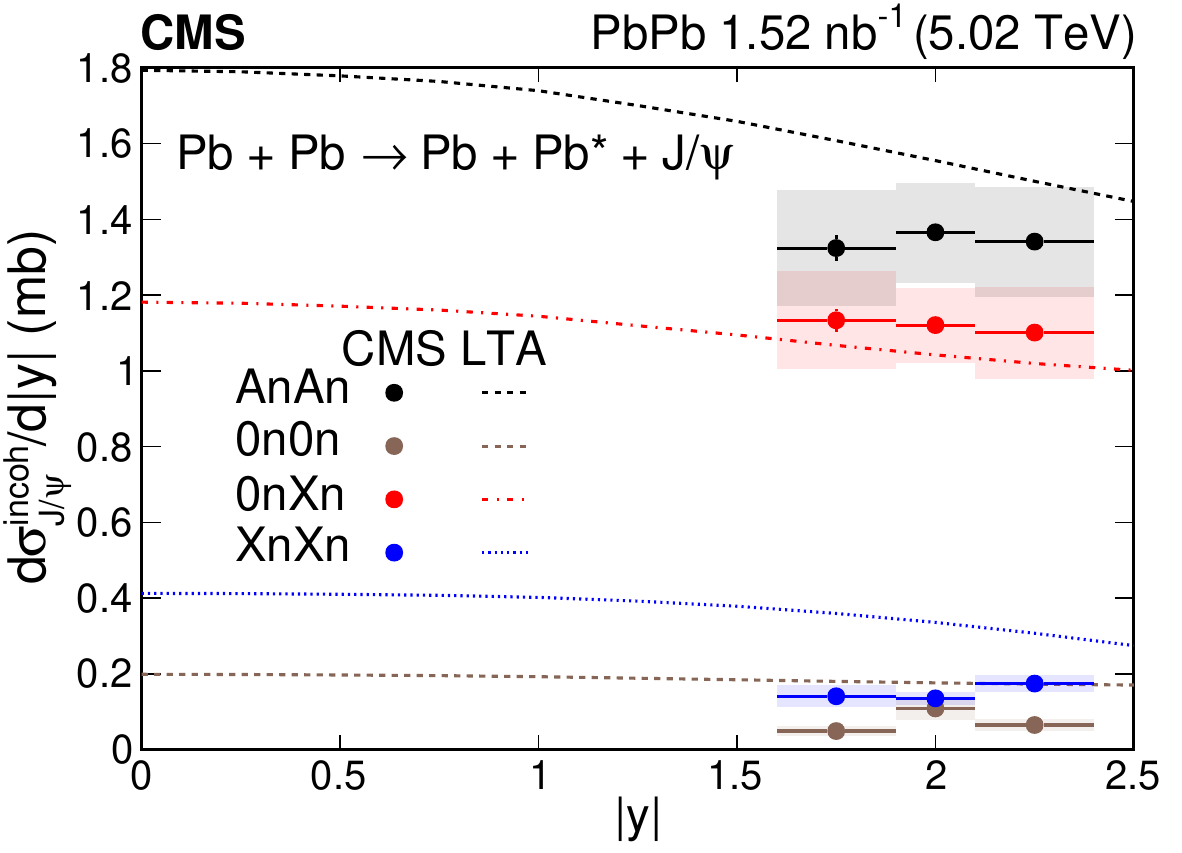}
\includegraphics[width=0.49\textwidth]{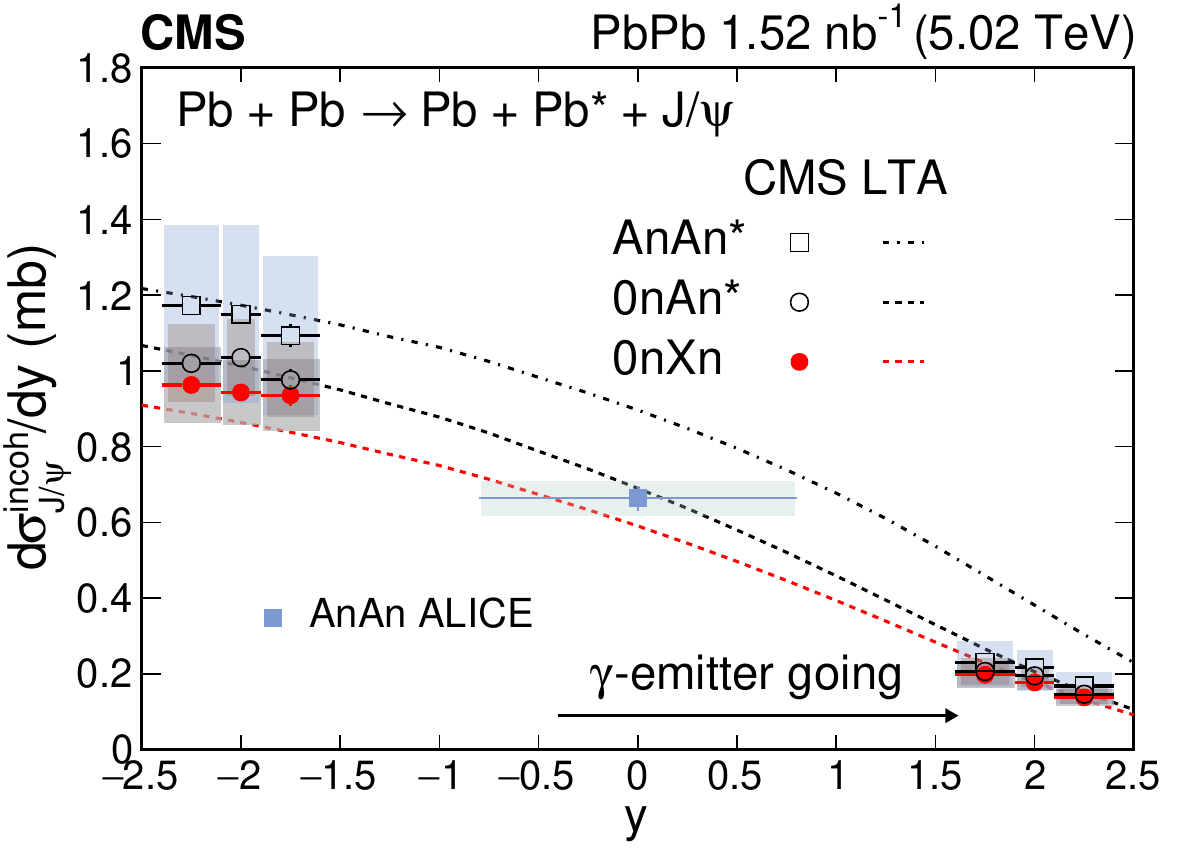}
\caption{
The total differential incoherent \PJGy photoproduction cross section versus $\abs{y}$ (left) and $y$ (right).
The right panel shows the results after disentangling the directional correlation between the produced \PJGy and emitted forward neutrons.
The small vertical bars and shaded boxes represent the statistical and systematic uncertainties, respectively. The horizontal bars show the bin widths.
Theoretical predictions from the LTA~\cite{Guzey:2018tlk,Kryshen:2023bxy} are shown by curves.
}
\label{fig:DSigmaDy}
\end{figure*}

The measured total differential incoherent $\PJGy$ photoproduction cross sections are reported in Fig.~\ref{fig:DSigmaDy} (left) in three rapidity bins within $1.6<\abs{y}<2.4$.
Results resolving the two-way ambiguity are shown versus signed $y$ in Fig.~\ref{fig:DSigmaDy} (right). Values at $-y$ and $+y$ reflect $\PJGy$ and neutrons with same or opposite rapidity signs, respectively.
The significant asymmetry between $-y$ and $+y$ confirms a strong correlation between the incoherent $\PJGy$ and neutrons. 
The higher values at $-y$ compared to $+y$ primarily result from the larger number of low energy photons.
The 0nAn* (AnAn*) categories are the sums of the 0nXn and 0n0n (0nXn, 0n0n, and XnXn) categories.
The ALICE data~\cite{ALICE:2023gcs}, although measured within $0.2<\pt<1.0\GeV$, are in good agreement with the rapidity distribution of CMS AnAn* data.
A model focusing on leading-twist contributions, which includes weak nuclear shadowing effects from multinucleon interference during the scattering process within the perturbative QCD framework (referred to as LTA~\cite{Guzey:2018tlk,Kryshen:2023bxy}), cannot describe the data in different neutron categories simultaneously, especially in the positive $y$ region.

\begin{figure}[hbt!]
\centering
\includegraphics[width=\linewidth]{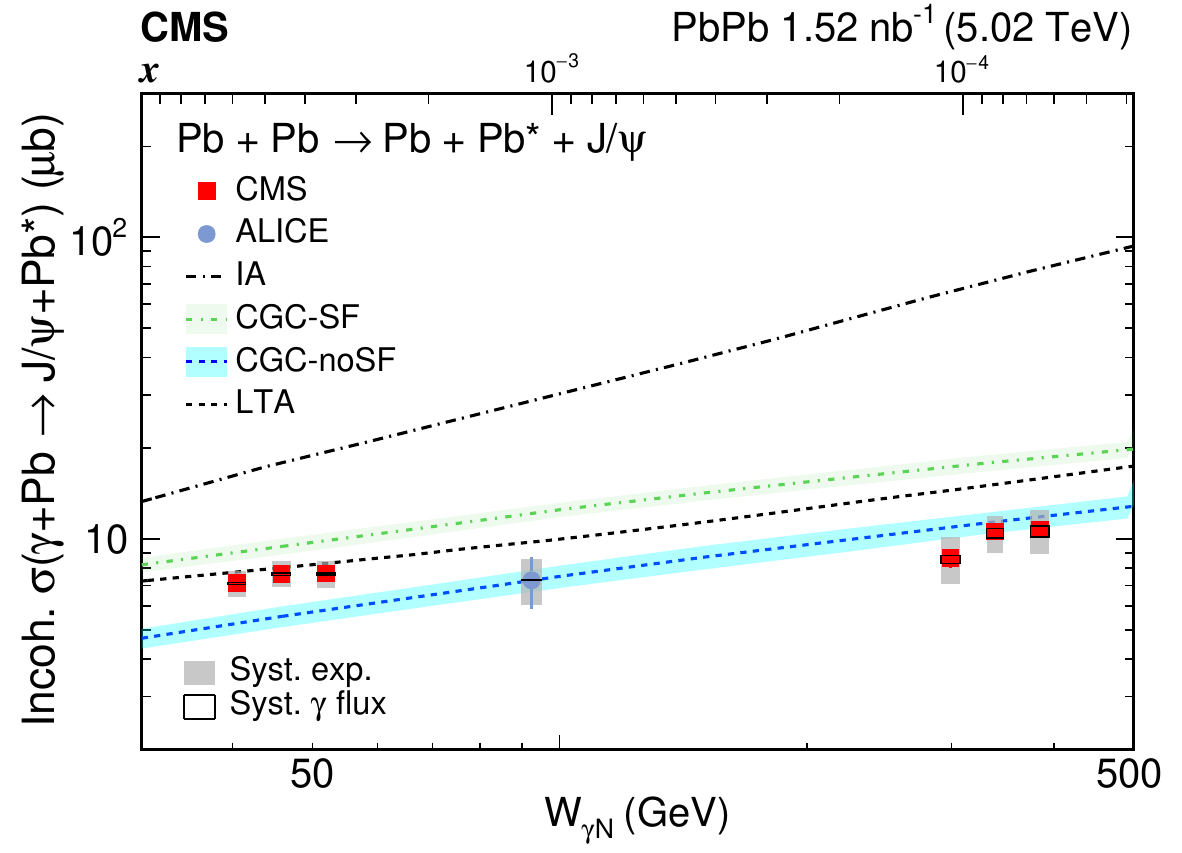}
\caption{
   The incoherent $\PJGy$ photoproduction cross section per \PGg{}\pb interaction as a function of \WgN (lower axis) or $x$ (upper axis) from the CMS measurement.
    The ALICE data ($\abs{y}<0.9$) measured at $\sqrtsNN = 2.76\TeV$~\cite{ALICE:2013wjo} is also displayed.
	The small vertical bars, shaded and open boxes represent the statistical, experimental, and theoretical (photon flux) uncertainties, respectively.
	Theoretical predictions from IA and LTA~\cite{Guzey:2018tlk,Kryshen:2023bxy}, and CGC~\cite{Mantysaari:2023xcu,Mantysaari:2022sux} models are shown by the curves, where the shaded bands are the theoretical uncertainties.
}
\label{fig:SigmaVsW}
\end{figure}

Following Eq.~(\ref{eq:sigma_0nAn}), the total incoherent $\PJGy$ photoproduction cross section (0nAn*) is corrected by the photon flux values to obtain the cross section per \PGg{}\pb interaction.
The results are shown as a function of $\WgN$ in Fig.~\ref{fig:SigmaVsW}. 
Results are compared with several theoretical models: IA and LTA~\cite{Guzey:2018tlk,Kryshen:2023bxy}, and CGC~\cite{Mantysaari:2023xcu,Mantysaari:2022sux}. The CGC model is based on the color glass condensate (CGC) framework, an effective theory that describes the high-gluon density regime of QCD~\cite{Gelis:2010nm}.
The CGC calculations use an impact-parameter dependent saturation formalism~\cite{Kowalski:2003hm} and offer predictions in two scenarios: with and without additional nucleon substructure fluctuations~\cite{Mantysaari:2023xcu,Mantysaari:2022sux}, labeled CGC-SF and CGC-noSF, respectively.
The experimental data are significantly lower than the IA prediction for $40<\WgN<400\GeV$, indicating strong nuclear modification of incoherent $\PJGy$ photoproduction in \PGg{}\pb interactions.
The LTA model describes the CMS data for $\WgN<60\GeV$, but seems to systematically overestimate the data at higher $\WgN$.
While the CGC-noSF model describes the ALICE and CMS data for $\WgN>90\GeV$, it underestimates the CMS data for $\WgN<60\GeV$. The CGC-SF model systematically overestimates the data.

\begin{figure}[hbt!]
	\centering
	\includegraphics[width=\linewidth]{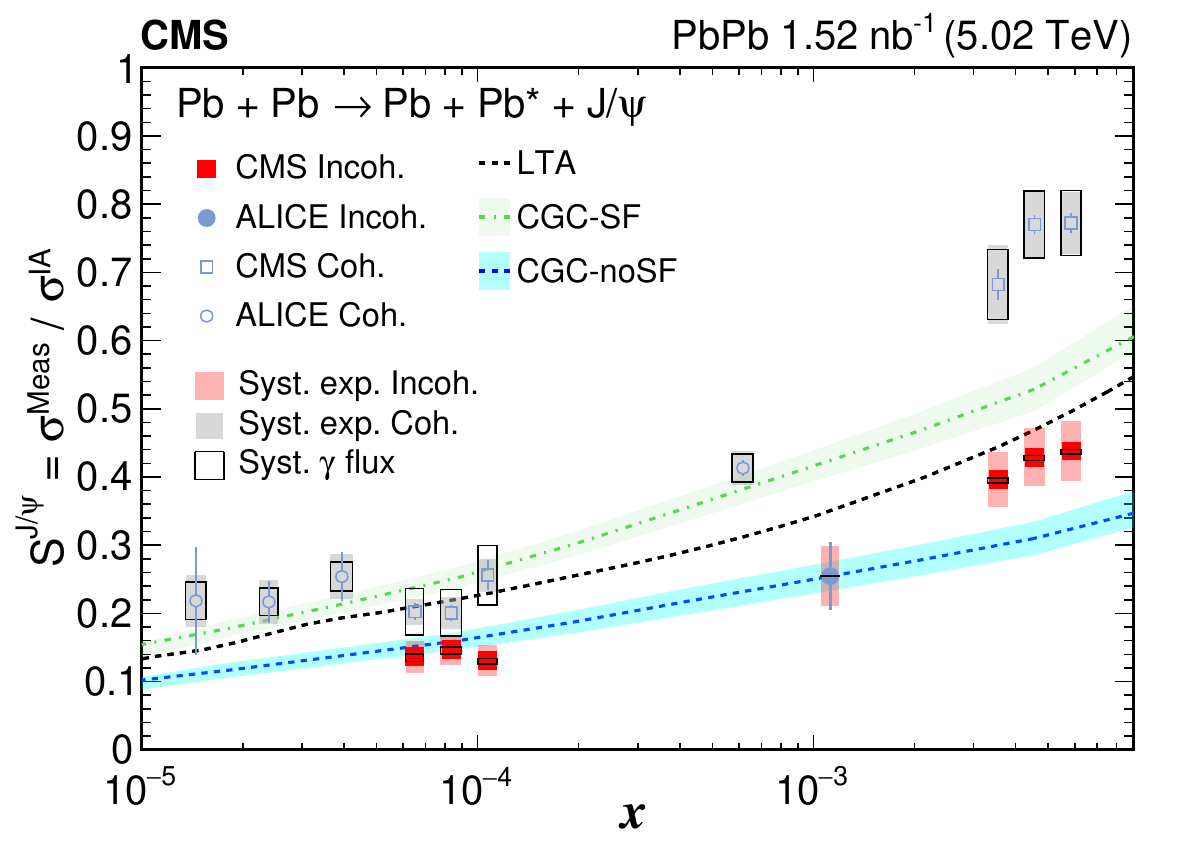}
	\caption{
	The nuclear suppression factor $\Sjpsi$ of incoherent (from this study and ALICE~\cite{ALICE:2013wjo}) and coherent (from CMS~\cite{CMS:2023snh} and ALICE~\cite{ALICE:2023jgu}) $\PJGy$ photoproduction as a function of $x$.
	The vertical bars and shaded and open boxes represent the statistical, experimental systematic, and theoretical systematic uncertainties, respectively.
	The prediction (incoherent) from the LTA~\cite{Guzey:2018tlk,Kryshen:2023bxy} and CGC~\cite{Mantysaari:2023xcu,Mantysaari:2022sux} models are shown by curves, where the shaded bands are the theoretical uncertainties.
	}
	\label{fig:suppression_vs_x}
\end{figure}

To quantify the impact of nuclear modifications on incoherent $\PJGy$ photoproduction, a nuclear suppression factor, $\Sjpsi$, is defined as a ratio of measured cross section to the IA prediction in Ref.~\cite{Guzey:2018tlk,Kryshen:2023bxy}.
The extracted $\Sjpsi$ values are plotted as a function of $x$ in Fig.~\ref{fig:suppression_vs_x}.
The suppression in the region of $x>3 \times 10^{-3}$ is approximately 0.4--0.5.
The \Sjpsi value drops rapidly to around 0.25 at $x{\approx}10^{-3}$ before plateauing at a value of around 0.14--0.18 for $x<10^{-4}$.
The $\Sjpsi$ of incoherent production is generally lower than that of coherent production for $x > 10^{-4}$. (Note that the nuclear gluon suppression factor of coherent $\PJGy$ photoproduction reported in Refs~\cite{CMS:2023snh,ALICE:2023jgu} is defined by $\RglPb=\sqrt{\Sjpsi}$).
However, at lower $x$, their $\Sjpsi$ values become more comparable, suggesting that nuclear suppression effects become insensitive to whether the process is coherent or incoherent at sufficiently small $x$.
Both LTA and CGC models fail to describe the observed $\Sjpsi$ distribution over the probed $x$ region.

\begin{figure}[hbt!]
	\centering
	\includegraphics[width=\linewidth]{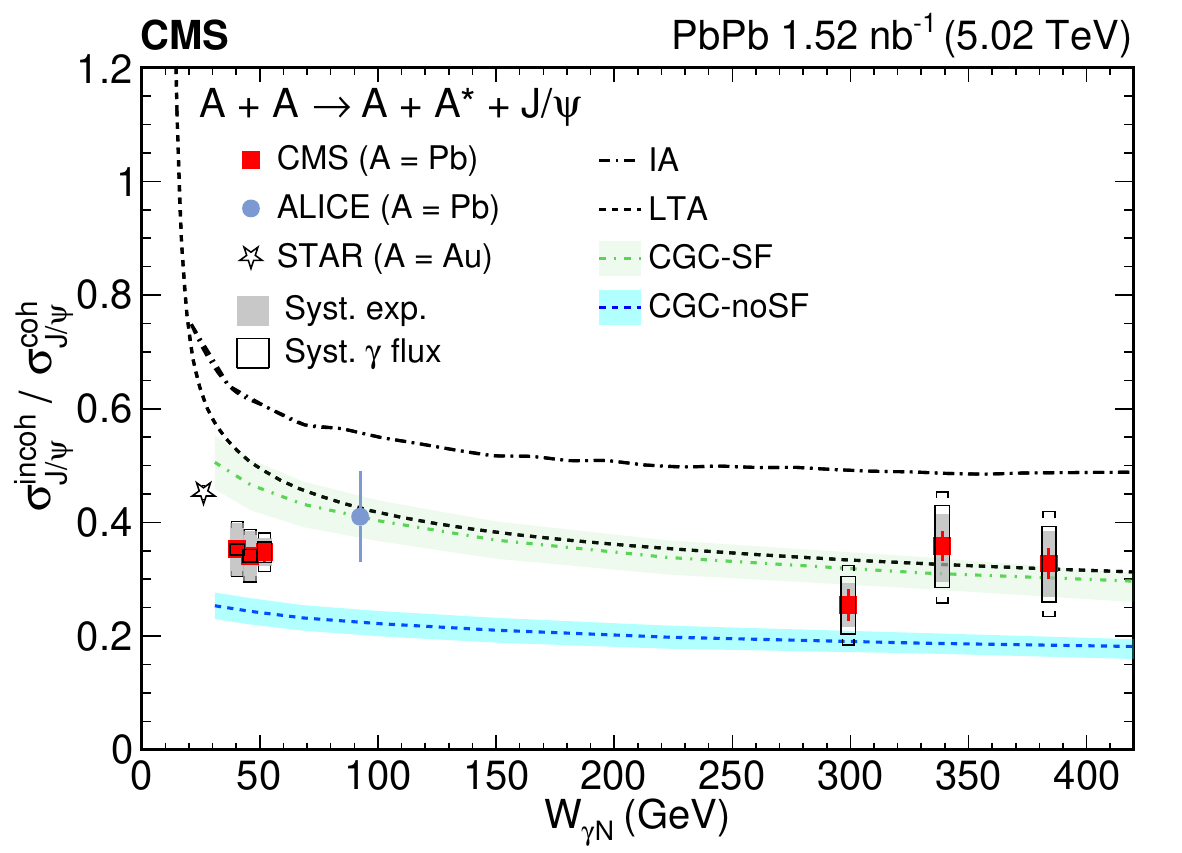}
	\caption{
	The ratio between incoherent and coherent $\PJGy$ photoproduction cross section as a function of \WgN.
	The midrapidity data from the STAR~\cite{STAR:2023nos} and ALICE~\cite{ALICE:2013wjo} experiments are also displayed.
	The vertical bars, the shaded boxes, open boxes, and brackets of the CMS data represent the statistical, experimental, theoretical (photon flux), and total uncertainties, respectively. 
	The vertical bar ALICE data represent the total uncertainties.
	The STAR error bars are invisible due to the assumption of substantial cancellation of systematic uncertainties.
	Theoretical predictions from the IA~\cite{Guzey:2013xba,Kryshen:2023bxy}, LTA~\cite{Guzey:2018tlk,Kryshen:2023bxy}, and CGC~\cite{Mantysaari:2023xcu,Mantysaari:2022sux} models are shown by curves, where the shaded bands are the theoretical uncertainties. 
	}
	\label{fig:RatioInCoh2Coh_vs_W}
\end{figure}

Figure~\ref{fig:RatioInCoh2Coh_vs_W} shows the ratio of measured incoherent to coherent $\PJGy$ cross sections, derived using the coherent results from Ref.~\cite{CMS:2023snh}.
The data are significantly lower than the IA prediction~\cite{Guzey:2013xba,Kryshen:2023bxy}, suggesting that the incoherent photoproduction is more strongly suppressed than the coherent photoproduction, likely due to underlying nuclear effects.
At $\WgN<60\GeV$, the LTA~\cite{Guzey:2018tlk,Kryshen:2023bxy} predicts a steeply rising trend toward lower energies, which is not supported by the data. This is despite the fact that the LTA model can successfully describe the incoherent $\PJGy$ cross section in this energy range, as shown in Fig.~\ref{fig:SigmaVsW}.
Although the CGC-SF prediction does not describe separately the coherent and incoherent cross sections well,  it qualitatively captures the overall trend of their ratio across the entire $\WgN$ range. In contrast, the CGC-noSF model significantly underestimates the data.
In the higher $\WgN$ region, both the LTA and CGC-SF models describe the data.
For the \WgN range probed, the CMS measured ratios stay around a constant value which agrees with previous measurements performed by the STAR~\cite{STAR:2023nos} and ALICE~\cite{ALICE:2013wjo} experiments. 
The nearly-flat ratio suggests that the recent results of slowly increasing coherent $\PJGy$ photoproduction cross section versus \WgN~\cite{CMS:2023snh,ALICE:2023jgu} should not be attributed to the aforementioned black disk limit~\cite{Frankfurt:2001nt}. If it were, the ratio would decrease at higher $\WgN$, as the nuclear inner structure becomes invisible.

In summary, this Letter presents the first energy-dependent measurements of incoherent $\PJGy$ photoproduction in ultraperipheral lead-lead collisions at a nucleon-nucleon center-of-mass energy of 5.02\TeV. 
The data covers a photon-nucleon energy range of 40--400\GeV, probing gluons $x$ values between $5.9 \times 10^{-3}$ and $6.5 \times 10^{-5}$. 
Compared to baseline predictions without nuclear effects, incoherent $\PJGy$ photoproduction cross sections exhibit greater suppression at lower $x$, and this suppression is more pronounced than that of coherent $\PJGy$ production.
The observed constant ratio between incoherent and coherent photoproduction cross sections suggests suppression mechanisms disfavoring simple black disk limit scenarios at small $x$. 
A model that includes gluon field fluctuations at the subnucleon level does not fully describe the incoherent and coherent cross sections, but captures the general trend of their ratio across the $\WgN$ range. 
This study offers important new insights into the evolution of fluctuating gluon fields within nuclei across a wide $x$ range, indicating the need for advancements in current theoretical models.
These measurements also provide valuable input for the planning of future facilities, such as the EIC~\cite{Accardi:2012qut} and LHeC~\cite{LHeC:2020van}, where photoproduction processes will be of great interest.

\begin{acknowledgments}
	We congratulate our colleagues in the CERN accelerator departments for the excellent performance of the LHC and thank the technical and administrative staffs at CERN and at other CMS institutes for their contributions to the success of the CMS effort. In addition, we gratefully acknowledge the computing centers and personnel of the Worldwide LHC Computing Grid and other centers for delivering so effectively the computing infrastructure essential to our analyses. Finally, we acknowledge the enduring support for the construction and operation of the LHC, the CMS detector, and the supporting computing infrastructure provided by the following funding agencies: SC (Armenia), BMBWF and FWF (Austria); FNRS and FWO (Belgium); CNPq, CAPES, FAPERJ, FAPERGS, and FAPESP (Brazil); MES and BNSF (Bulgaria); CERN; CAS, MoST, and NSFC (China); MINCIENCIAS (Colombia); MSES and CSF (Croatia); RIF (Cyprus); SENESCYT (Ecuador); ERC PRG, RVTT3 and MoER TK202 (Estonia); Academy of Finland, MEC, and HIP (Finland); CEA and CNRS/IN2P3 (France); SRNSF (Georgia); BMBF, DFG, and HGF (Germany); GSRI (Greece); NKFIH (Hungary); DAE and DST (India); IPM (Iran); SFI (Ireland); INFN (Italy); MSIP and NRF (Republic of Korea); MES (Latvia); LMTLT (Lithuania); MOE and UM (Malaysia); BUAP, CINVESTAV, CONACYT, LNS, SEP, and UASLP-FAI (Mexico); MOS (Montenegro); MBIE (New Zealand); PAEC (Pakistan); MES and NSC (Poland); FCT (Portugal); MESTD (Serbia); MICIU/AEI and PCTI (Spain); MOSTR (Sri Lanka); Swiss Funding Agencies (Switzerland); MST (Taipei); MHESI and NSTDA (Thailand); TUBITAK and TENMAK (Turkey); NASU (Ukraine); STFC (United Kingdom); DOE and NSF (USA).
\end{acknowledgments}
\bibliography{auto_generated}

\appendix
\section{Method of incoherent \texorpdfstring{\PJGy}{J/psi} signal yield extraction \label{appendix:a} }

The reconstructed dimuon candidates can originate from several sources: coherent and incoherent $\PJGy$ photoproduction; $\PJGy$ coming from the decay of coherent and incoherent photoproduced $\PGyP{2S}$ (referred to as ``feed-down''); and the dimuon QED continuum resulting from the $\PGg\PGg \to \PGmp\PGmm$ process. 
To extract the incoherent $\PJGy$ photoproduction contribution, a two-step analysis is performed by first fitting the uncorrected dimuon invariant mass distribution in the coherent-process-dominated kinematic region of $\pt<0.2\GeV$, and then fitting the corrected dimuon \pt spectrum within the $\PJGy$ mass window of $2.95<m_{\mumu}<3.25\GeV$. 
In the first step, two Crystal Ball functions are used to describe the $\PJGy$ and $\PGyP{2S}$ signals, while a third-order polynomial function parametrizes the QED background ($\PGg\PGg \to \mumu$), following the fitting method described in Ref.~\cite{CMS:2023snh}. The contributions from the QED and feed-down processes are determined in this step. In the second step, by taking advantage of the different \pt spectra between coherent and incoherent $\PJGy$ photoproduction, in addition to constraints on the contributions from the QED continuum and the feed-down, the incoherent $\PJGy$ photoproduction yield is calculated using fitting techniques similar to those of Refs.~\cite{Acharya:2019vlb,ALICE:2021gpt,ALICE:2023jgu,CMS:2023snh}. 

Considering the long tail of the QED continuum in the high-$\pt$ region, which may arise from high-order $\gamma\gamma$ processes or semi-coherent processes as discussed in Refs.~\cite{CMS:2020skx,Baur:2007zz,Klein:2018fmp}, the $\pt$ shape from the sideband region of the $\PJGy$ mass peak is used. While the \pt spectrum in the sideband is an averaged $\pt$ shape of the lower-mass sideband ($2.75<m_{\mumu}<2.90\GeV$) and higher-mass sideband ($3.30<m_{\mumu}<3.35\GeV$), the lower or higher sideband alone is used for estimating the systematic uncertainties.
Previous studies~\cite{Acharya:2019vlb,ALICE:2021gpt,ALICE:2023jgu,CMS:2023snh} neglected the contribution of $\PJGy$ from direct incoherent photoproduced $\PGyP{2S}$ decays with nucleon dissociation. 
In this study, we obtained the \pt template for this contribution by assuming that the direct incoherent photoproduced $\PGyP{2S}$ has the same $\pt^2$ slope as measured by the H1 experiment at HERA~\cite{H1:2002yab}. 
The \pt shape of direct incoherent photoproduced $\PJGy$ with nucleon dissociation is modeled using an empirical function, employed by the H1~\cite{H1:2013okq}, ALICE~\cite{Acharya:2019vlb,ALICE:2021gpt,ALICE:2023jgu}, and CMS~\cite{CMS:2023snh} experiments: $\rd{N}/\rd{\pt} \approx \pt [1+(b_{\text{pd}}/n_{\text{pd}}) \pt^{2}]^{-n_{\text{pd}}}$. The two fit parameters, $b_{\text{pd}}$ and $n_{\text{pd}}$ are fixed to the values obtained by the H1 experiment~\cite{H1:2013okq} as default, but are treated as free parameters in the fit for systematic uncertainty estimation.
Normalizations of coherent and incoherent $\PGyP{2S}$ feed-down to $\PJGy$ are constrained to those of the prompt coherent and incoherent $\PJGy$ photoproduction components according to the cross section ratios of $\PGyP{2S}$ to $\PJGy$ extracted from the fit to the invariant mass distributions within the coherent-process-dominant \pt ($<0.2\GeV$) region, following Eqs.~(A2-A4) in Ref.~\cite{CMS:2023snh}. The primary fitting procedures are consistent with the well-established methods in Refs.~\cite{Acharya:2019vlb,ALICE:2021gpt,ALICE:2023jgu}.
Since the incoherent $\PJGy$ has a broad $\pt$ distribution, directly extracting the total uncorrected yield from the raw $\pt$ distribution and then correcting it by its reconstruction efficiency would require calculating the $\pt$-integrated efficiency. This would depend on the input $\pt$ shape. To mitigate the uncertainty regarding the $\pt$-integrated efficiency arising from the input $\pt$ shape, a $\pt$-dependent efficiency is applied on the uncorrected $\pt$ spectra. Subsequently, a fit of these corrected $\pt$ spectra allows the direct extraction of the corrected total incoherent $\PJGy$ yield $N^{\text{incoh}}_{\PJGy}/(A \epsilon)_{\PJGy}$.

\begin{figure*}[hbt!]
	\centering
	\includegraphics[width=0.49\textwidth]{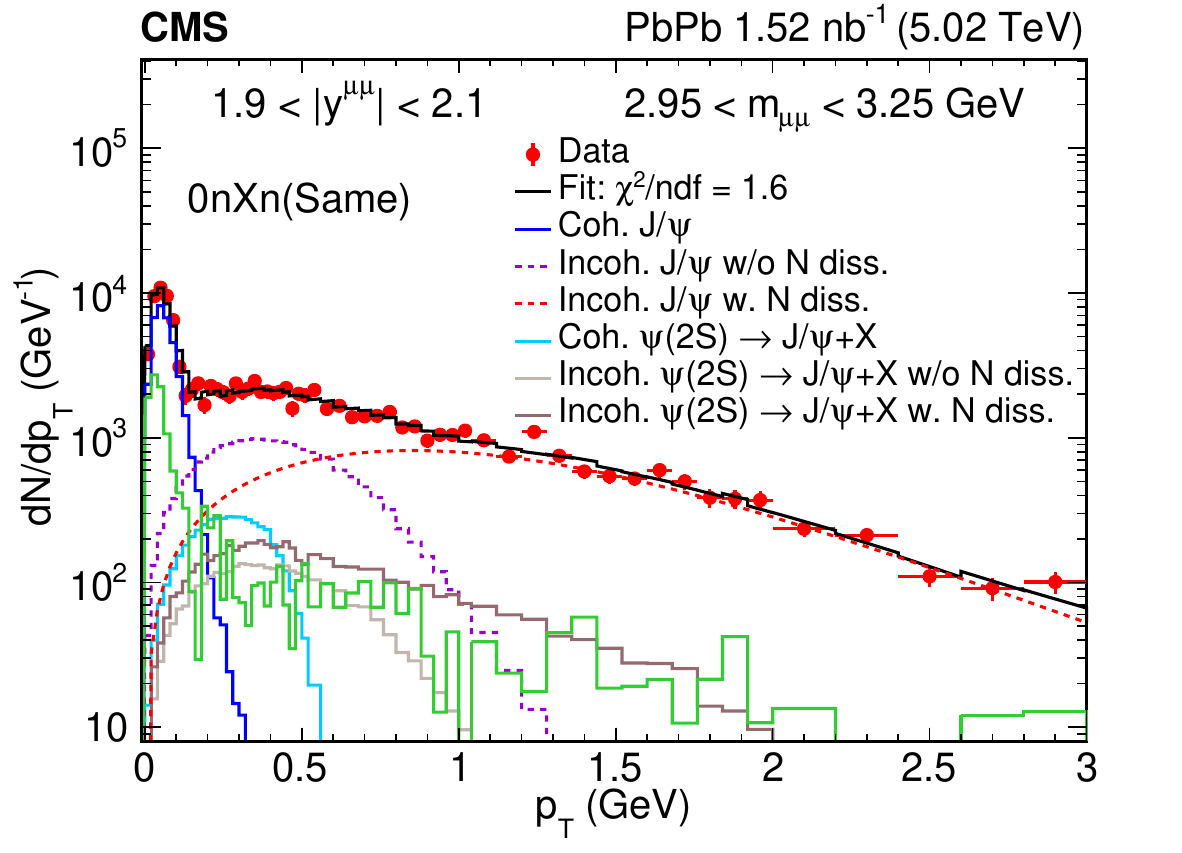}
	\includegraphics[width=0.49\textwidth]{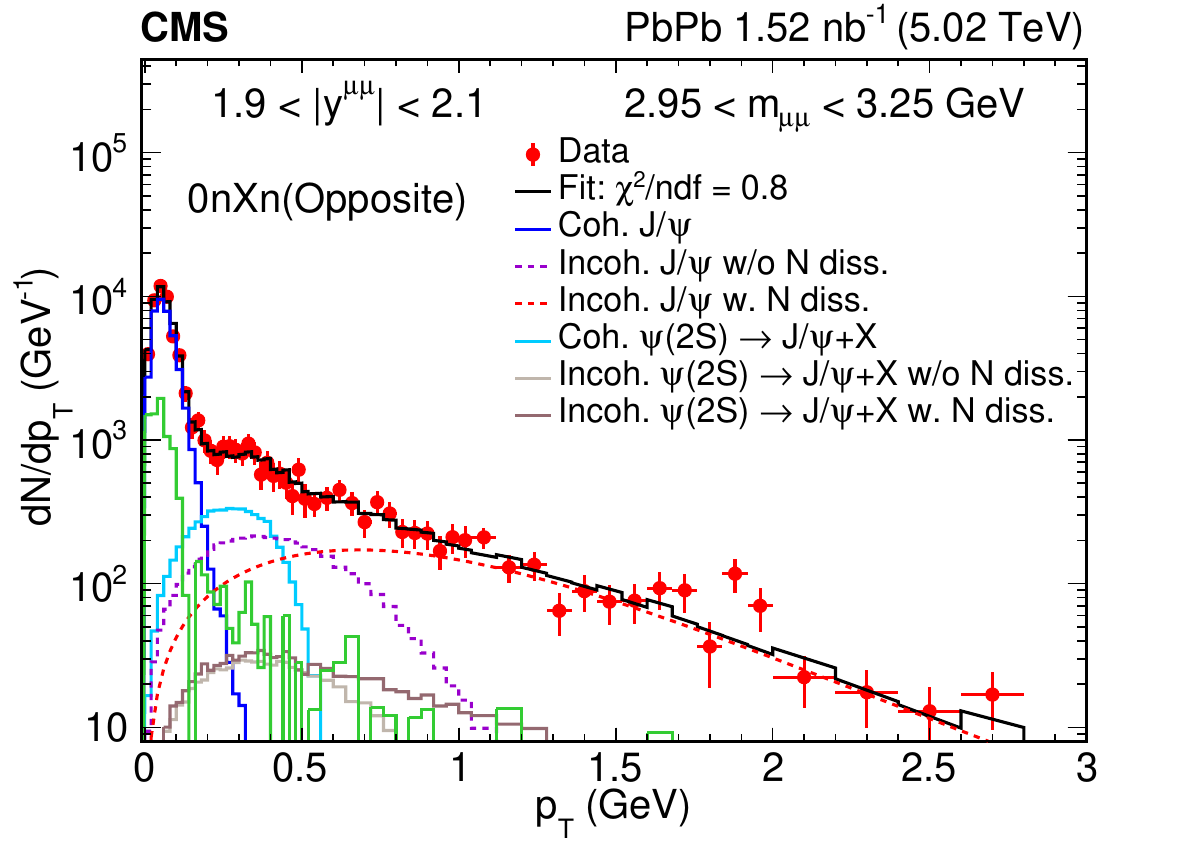}
	\caption{
	The corrected transverse momentum spectra of $\mumu$ pairs with a dimuon invariant mass in the $\PJGy$ mass window of ($2.95<m_{\mumu}<3.25\GeV$) in ``0nXn(Same)" (left) and ``0nXn(Opposite)" (right) events.
	The results of the fit are shown by the various curves. 
	The vertical bars on the data points represent the statistical uncertainty.
	}
	\label{fig:SignalsFitting}
\end{figure*}

Figure~\ref{fig:SignalsFitting} shows examples of the corrected dimuon \pt spectrum for dimuon candidates having an invariant mass in the $\PJGy$ mass window.
The cases where $\PJGy$ candidates and neutrons have the same or opposite rapidity signs are denoted as ``$\PJGy$-Xn(Same)" and ``$\PJGy$-Xn(Opposite)".
The magnitudes of coherent contributions in the two panels are similar, whereas the incoherent contribution in the left panel is significantly higher than that in the right panel. 
This behavior confirms the theoretical expectations~\cite{Guzey:2013jaa,Guzey:2016piu} that the emitted neutrons in coherent $\PJGy$ production events are primarily caused by additional soft-photon exchange between the two nuclei, resulting in no correlation between the neutron direction and the $\PJGy$ rapidity. In contrast, the neutrons emitted from incoherent $\PJGy$ production exhibit a strong correlation between the neutron directions and the $\PJGy$ rapidity, as these neutrons are mainly induced by the incoherent interaction process itself. Similar findings are also reported in a recent STAR measurement~\cite{STAR:2023vvb}.

\cleardoublepage \section{The CMS Collaboration \label{app:collab}}\begin{sloppypar}\hyphenpenalty=5000\widowpenalty=500\clubpenalty=5000
\cmsinstitute{Yerevan Physics Institute, Yerevan, Armenia}
{\tolerance=6000
V.~Chekhovsky, A.~Hayrapetyan, V.~Makarenko\cmsorcid{0000-0002-8406-8605}, A.~Tumasyan\cmsAuthorMark{1}\cmsorcid{0009-0000-0684-6742}
\par}
\cmsinstitute{Institut f\"{u}r Hochenergiephysik, Vienna, Austria}
{\tolerance=6000
W.~Adam\cmsorcid{0000-0001-9099-4341}, J.W.~Andrejkovic, L.~Benato\cmsorcid{0000-0001-5135-7489}, T.~Bergauer\cmsorcid{0000-0002-5786-0293}, S.~Chatterjee\cmsorcid{0000-0003-2660-0349}, K.~Damanakis\cmsorcid{0000-0001-5389-2872}, M.~Dragicevic\cmsorcid{0000-0003-1967-6783}, P.S.~Hussain\cmsorcid{0000-0002-4825-5278}, M.~Jeitler\cmsAuthorMark{2}\cmsorcid{0000-0002-5141-9560}, N.~Krammer\cmsorcid{0000-0002-0548-0985}, A.~Li\cmsorcid{0000-0002-4547-116X}, D.~Liko\cmsorcid{0000-0002-3380-473X}, I.~Mikulec\cmsorcid{0000-0003-0385-2746}, J.~Schieck\cmsAuthorMark{2}\cmsorcid{0000-0002-1058-8093}, R.~Sch\"{o}fbeck\cmsAuthorMark{2}\cmsorcid{0000-0002-2332-8784}, D.~Schwarz\cmsorcid{0000-0002-3821-7331}, M.~Sonawane\cmsorcid{0000-0003-0510-7010}, W.~Waltenberger\cmsorcid{0000-0002-6215-7228}, C.-E.~Wulz\cmsAuthorMark{2}\cmsorcid{0000-0001-9226-5812}
% ...
\par}
\vskip\cmsinstskip
\dag:~Deceased\\
$^{1}$Also at Yerevan State University, Yerevan, Armenia\\
$^{2}$Also at TU Wien, Vienna, Austria\\
$^{3}$Also at Ghent University, Ghent, Belgium\\
$^{4}$Also at Universidade do Estado do Rio de Janeiro, Rio de Janeiro, Brazil\\
$^{5}$Also at FACAMP - Faculdades de Campinas, Sao Paulo, Brazil\\
$^{6}$Also at Universidade Estadual de Campinas, Campinas, Brazil\\
$^{7}$Also at Federal University of Rio Grande do Sul, Porto Alegre, Brazil\\
$^{8}$Also at University of Chinese Academy of Sciences, Beijing, China\\
$^{9}$Also at China Center of Advanced Science and Technology, Beijing, China\\
$^{10}$Also at University of Chinese Academy of Sciences, Beijing, China\\
$^{11}$Also at China Spallation Neutron Source, Guangdong, China\\
$^{12}$Now at Henan Normal University, Xinxiang, China\\
$^{13}$Also at University of Shanghai for Science and Technology, Shanghai, China\\
$^{14}$Now at The University of Iowa, Iowa City, Iowa, USA\\
$^{15}$Also at an institute formerly covered by a cooperation agreement with CERN\\
$^{16}$Also at Helwan University, Cairo, Egypt\\
$^{17}$Also at Suez University, Suez, Egypt\\
$^{18}$Now at British University in Egypt, Cairo, Egypt\\
$^{19}$Also at Purdue University, West Lafayette, Indiana, USA\\
$^{20}$Also at Universit\'{e} de Haute Alsace, Mulhouse, France\\
$^{21}$Also at Istinye University, Istanbul, Turkey\\
$^{22}$Also at Tbilisi State University, Tbilisi, Georgia\\
$^{23}$Also at an international laboratory covered by a cooperation agreement with CERN\\
$^{24}$Also at The University of the State of Amazonas, Manaus, Brazil\\
$^{25}$Also at University of Hamburg, Hamburg, Germany\\
$^{26}$Also at RWTH Aachen University, III. Physikalisches Institut A, Aachen, Germany\\
$^{27}$Also at Bergische University Wuppertal (BUW), Wuppertal, Germany\\
$^{28}$Also at Brandenburg University of Technology, Cottbus, Germany\\
$^{29}$Also at Forschungszentrum J\"{u}lich, Juelich, Germany\\
$^{30}$Also at CERN, European Organization for Nuclear Research, Geneva, Switzerland\\
$^{31}$Also at HUN-REN ATOMKI - Institute of Nuclear Research, Debrecen, Hungary\\
$^{32}$Now at Universitatea Babes-Bolyai - Facultatea de Fizica, Cluj-Napoca, Romania\\
$^{33}$Also at MTA-ELTE Lend\"{u}let CMS Particle and Nuclear Physics Group, E\"{o}tv\"{o}s Lor\'{a}nd University, Budapest, Hungary\\
$^{34}$Also at HUN-REN Wigner Research Centre for Physics, Budapest, Hungary\\
$^{35}$Also at Physics Department, Faculty of Science, Assiut University, Assiut, Egypt\\
$^{36}$Also at Punjab Agricultural University, Ludhiana, India\\
$^{37}$Also at University of Visva-Bharati, Santiniketan, India\\
$^{38}$Also at Indian Institute of Science (IISc), Bangalore, India\\
$^{39}$Also at Amity University Uttar Pradesh, Noida, India\\
$^{40}$Also at IIT Bhubaneswar, Bhubaneswar, India\\
$^{41}$Also at Institute of Physics, Bhubaneswar, India\\
$^{42}$Also at University of Hyderabad, Hyderabad, India\\
$^{43}$Also at Deutsches Elektronen-Synchrotron, Hamburg, Germany\\
$^{44}$Also at Isfahan University of Technology, Isfahan, Iran\\
$^{45}$Also at Sharif University of Technology, Tehran, Iran\\
$^{46}$Also at Department of Physics, University of Science and Technology of Mazandaran, Behshahr, Iran\\
$^{47}$Also at Department of Physics, Faculty of Science, Arak University, ARAK, Iran\\
$^{48}$Also at Italian National Agency for New Technologies, Energy and Sustainable Economic Development, Bologna, Italy\\
$^{49}$Also at Centro Siciliano di Fisica Nucleare e di Struttura Della Materia, Catania, Italy\\
$^{50}$Also at Universit\`{a} degli Studi Guglielmo Marconi, Roma, Italy\\
$^{51}$Also at Scuola Superiore Meridionale, Universit\`{a} di Napoli 'Federico II', Napoli, Italy\\
$^{52}$Also at Fermi National Accelerator Laboratory, Batavia, Illinois, USA\\
$^{53}$Also at Lulea University of Technology, Lulea, Sweden\\
$^{54}$Also at Consiglio Nazionale delle Ricerche - Istituto Officina dei Materiali, Perugia, Italy\\
$^{55}$Also at Institut de Physique des 2 Infinis de Lyon (IP2I ), Villeurbanne, France\\
$^{56}$Also at Department of Applied Physics, Faculty of Science and Technology, Universiti Kebangsaan Malaysia, Bangi, Malaysia\\
$^{57}$Also at Consejo Nacional de Ciencia y Tecnolog\'{i}a, Mexico City, Mexico\\
$^{58}$Also at Trincomalee Campus, Eastern University, Sri Lanka, Nilaveli, Sri Lanka\\
$^{59}$Also at Saegis Campus, Nugegoda, Sri Lanka\\
$^{60}$Also at National and Kapodistrian University of Athens, Athens, Greece\\
$^{61}$Also at Ecole Polytechnique F\'{e}d\'{e}rale Lausanne, Lausanne, Switzerland\\
$^{62}$Also at Universit\"{a}t Z\"{u}rich, Zurich, Switzerland\\
$^{63}$Also at Stefan Meyer Institute for Subatomic Physics, Vienna, Austria\\
$^{64}$Also at Laboratoire d'Annecy-le-Vieux de Physique des Particules, IN2P3-CNRS, Annecy-le-Vieux, France\\
$^{65}$Also at Near East University, Research Center of Experimental Health Science, Mersin, Turkey\\
$^{66}$Also at Konya Technical University, Konya, Turkey\\
$^{67}$Also at Izmir Bakircay University, Izmir, Turkey\\
$^{68}$Also at Adiyaman University, Adiyaman, Turkey\\
$^{69}$Also at Bozok Universitetesi Rekt\"{o}rl\"{u}g\"{u}, Yozgat, Turkey\\
$^{70}$Also at Marmara University, Istanbul, Turkey\\
$^{71}$Also at Milli Savunma University, Istanbul, Turkey\\
$^{72}$Also at Kafkas University, Kars, Turkey\\
$^{73}$Now at Istanbul Okan University, Istanbul, Turkey\\
$^{74}$Also at Hacettepe University, Ankara, Turkey\\
$^{75}$Also at Erzincan Binali Yildirim University, Erzincan, Turkey\\
$^{76}$Also at Istanbul University -  Cerrahpasa, Faculty of Engineering, Istanbul, Turkey\\
$^{77}$Also at Yildiz Technical University, Istanbul, Turkey\\
$^{78}$Also at School of Physics and Astronomy, University of Southampton, Southampton, United Kingdom\\
$^{79}$Also at IPPP Durham University, Durham, United Kingdom\\
$^{80}$Also at Monash University, Faculty of Science, Clayton, Australia\\
$^{81}$Also at Universit\`{a} di Torino, Torino, Italy\\
$^{82}$Also at Bethel University, St. Paul, Minnesota, USA\\
$^{83}$Also at Karamano\u {g}lu Mehmetbey University, Karaman, Turkey\\
$^{84}$Also at California Institute of Technology, Pasadena, California, USA\\
$^{85}$Also at United States Naval Academy, Annapolis, Maryland, USA\\
$^{86}$Also at Ain Shams University, Cairo, Egypt\\
$^{87}$Also at Bingol University, Bingol, Turkey\\
$^{88}$Also at Georgian Technical University, Tbilisi, Georgia\\
$^{89}$Also at Sinop University, Sinop, Turkey\\
$^{90}$Also at Erciyes University, Kayseri, Turkey\\
$^{91}$Also at Horia Hulubei National Institute of Physics and Nuclear Engineering (IFIN-HH), Bucharest, Romania\\
$^{92}$Now at another institute formerly covered by a cooperation agreement with CERN\\
$^{93}$Also at Texas A\&M University at Qatar, Doha, Qatar\\
$^{94}$Also at Kyungpook National University, Daegu, Korea\\
$^{95}$Also at Yerevan Physics Institute, Yerevan, Armenia\\
$^{96}$Also at another international laboratory covered by a cooperation agreement with CERN\\
$^{97}$Also at Imperial College, London, United Kingdom\\
$^{98}$Also at Institute of Nuclear Physics of the Uzbekistan Academy of Sciences, Tashkent, Uzbekistan\\
$^{99}$Also at another institute formerly covered by a cooperation agreement with CERN\\
\end{sloppypar}
\end{document}